\documentclass[preprint,preprintnumbers,amsmath,amssymb,nofootinbib]{revtex4-1}
\pdfoutput=1
\usepackage{bbm, amsfonts, booktabs, mathrsfs, epsfig, graphics, dcolumn, bm, amsmath,color}
\usepackage[colorlinks,citecolor=blue,urlcolor=blue,linkcolor=blue]{hyperref}
\usepackage{multirow}

\begin{document}

\title{TeV SUSY dark matter confronted with the current direct and indirect detection data}

\author{
Murat Abdughani$^{1,2}$,  
Jie Ren$^{1,2}$,         
Jun Zhao$^{1,2,3}$,
}
\affiliation{
$^1$ CAS Key Laboratory of Theoretical Physics, Institute of Theoretical Physics, Chinese Academy of Sciences, Beijing 100190, China\\
$^2$ School of Physical Sciences, University of Chinese Academy of Sciences, Beijing 100049, China \\
$^3$ Institute of Theoretical Physics, College of Applied Science, Beijing University of Technology, Beijing 100124, China
}

\begin{abstract}
In the minimal supersymmetric standard model (MSSM)
the lightest superparticle (LSP) can be a TeV neutralino
(mainly dominated by higgsino or wino) which
serves as a dark matter candidate with correct thermal relic density.
In this work we confront the 1-2 TeV neutralino dark matter with the latest
direct and indirect detections from PandaX and AMS-02/DAMPE.
Considering various scenarios
with decoupled sfermions, with \textit{A}-mediated annihilation, with squark or
stop coannihilation,
we find that the parameter space is stringently constrained by the direct detection limits.
In the allowed parameter space, the TeV neutralino dark matter
annihilation contribution to the anti-proton flux is found to agree with the AMS-02 data
while its contribution to eletron/positron flux is too small to cause any visible excess.
The current survived parameter space can be mostly covered by the future direct detection experiment
LZ7.2T.

\end{abstract}

\maketitle


\section{Introduction}

Identifying the nature of the cosmic dark matter is a primary topic in today's particle physics and
 cosmology. In the popular minimal supersymmetric standard model (MSSM), the lightest neutralino is
a natural candidate for the cosmic cold dark matter.
In general, a neutralino dark matter with a mass around 100 GeV (a typical WIMP) can be effectively
explored at direct detection experiments and the LHC. However, a neutralino dark matter above TeV scale
or below GeV scale is hard to detect at the LHC. On the other hand,
the current direct detections \cite{Cui:2017nnn} have relatively low sensitivities to
such a super-heavy or ultra-light neutralino dark matter and hence the limits on their interactions
with the nucleon are rather weak.
For the TeV scale dark matter, another motivation comes from the recent DAMPE
observation \cite{Ambrosi:2017wek}
of a plausible electron/positron excess at TeV energy which may indicate a heavy dark matter at TeV scale.
Theoretically, in the MSSM an ultra-light GeV scale neutralino
dark matter can only be achieved in some unnatural limits (say the alignment limit without
decoupling \cite{Duan:2017ucw,Ren:2017ymm}),
but a TeV scale neutralino dark matter can be naturally obtained with correct
thermal relic density.
So a TeV scale neutralino dark matter is an interesting scenario to study.

Such a TeV neutralino dark matter has been discussed
in the literature \cite{Beneke:2016ync, Beneke:2016jpw, Abdughani:2017dqs, Bramante:2015una, Duan:2018rls,Han:2016xet,Kobakhidze:2016mdx,Wu:2017kgr,Han:2016gvr,diCortona:2014yua,Kowalska:2018toh,Mitridate:2017izz}.
In this work we intend to give a more complete study by considering various
scenarios with decoupled sfermions, with \textit{A}-mediated annihilation, with squark or stop coannihilation.
Under the requirement of giving correct thermal relic density, we will show its components.
Then we will demonstrate the constraints of the latest direct detections on its parameter space.
In the allowed parameter space we will show the contributions
of its annihilation to the anti-proton and eletron/positron cosmic-ray fluxes, which will be compared with
the AMS-02 and DAMPE data.

The structure of this paper is organized as follows. In Section \ref{section2} we show the scenarios
and the components
of the TeV neutralino dark matter under the requirement of giving correct thermal relic density.
 In Section \ref{section3} we confront
the TeV neutralino dark matter in various scenarios with the latest direct detection limits
from PandaX and indirect constraints from the anti-proton
and eletron/positron cosmic-ray fluxes from AMS-02 and DAMPE data.
Finally, we draw our conclusions in Section \ref{section4}.

\section{TeV neutralino dark matter with correct relic density }\label{section2}
In the MSSM the two neutral higgsinos ($\tilde{H}_u^0$ and $\tilde{H}_d^0$) and the two neutral guaginos
($\tilde{B}$ and $\tilde{W}^0$) are mixed to form four mass eigenstates called neutralinos.
In the gauge-eigenstate basis ($ \tilde{B}, \tilde{W}^0, \tilde{H}_d, \tilde{H}_u $), the neutralinos
are defined as $ \tilde{\chi}_i^0 = Z_N^{ij}(\tilde{B}, \tilde{W}^0, \tilde{H}_d, \tilde{H}_u) $ while
the charginos are defined as $ \tilde{\chi}_i^\pm = Z_{\pm}^{ij}(\tilde W^{\pm}, \tilde H^{\pm}) $,
where $ \tilde{B}$, $\tilde{W}$, $\tilde{H}_d$, $\tilde{H}_u $ are respectively the bino, wino, and higgsino
fields, and $ Z_N^{ij} $ and $ Z_{\pm}^{ij} $ are neutralino and chargino mixing matrices.

In our analysis we take the bino mass $ M_1 $, wino mass $ M_2 $, and higgsino mass $ \mu $
in the range of 1-7 TeV
when we scan over the parameter space. We require the neutralino dark matter (the lightest neutralino)
in the range of 1-2 TeV. In our scan  we use MicrOEMGAs \cite{Belanger:2013oya} to calculate the thermal
relic density $\Omega_{\tilde{\chi}} h^2$ and the cross sections and require the neutralino dark matter to
provide the relic density in the $2\sigma $ range of the measured value \cite{Ade:2013zuv}.
We fix $ tan\beta = 30 $ and trilinear terms $ A_i = 0 $.
We consider the following scenarios:
\begin{itemize}
\item[(i)] \textbf{Decoupled Case}:
Motivated from the split supersymmetry \cite{Wells:2003tf}, in this case we set sfermion mass
parameters and the CP-odd Higgs mass as heavy as 10 TeV to decouple them from gauginos.
In this case the main processes which affect the dark matter relic density and annihilation
cross section involve the interactions between neutralinos, charginos and the $W/Z$ or Higgs boson.
The interactions of gauge boson with neutralino and chargino are given by
(we used the conventions in \cite{Rosiek:1995kg} for particles, couplings,
and their diagonalization matrices)
\begin{eqnarray}
&& {\frac{e}{s_W}} \bar{\chi}_j \gamma^{\mu} \left[(Z_N^{2i}
Z_+^{1j\star} - {\frac{1}{\sqrt{2}}} Z_N^{4i} Z_+^{2j\star}) P_L +
(Z_N^{2i\star} Z_-^{1j} + {\frac{1}{\sqrt{2}}} Z_N^{3i\star} Z_-^{2j})
P_R\right] \chi^0_i W^+_{\mu}  \nonumber\\
&& - {\frac{e}{2s_Wc_W}} \bar{\chi}_i \gamma^{\mu} \left(Z_+^{1i\star}
Z_+^{1j} P_L + Z_-^{1i} Z_-^{1j\star} P_R + (c_W^2 - s_W^2)
\delta^{ij} \right) \chi_j Z_{\mu} \nonumber\\
&& + {\frac{e}{4 s_W c_W}} \bar{\chi}_i^0 {\gamma}^{\mu} \left(
(Z_N^{4i\star} Z_N^{4j} - Z_N^{3i\star} Z_N^{3j}) P_L - (Z_N^{4i}
Z_N^{4j\star} - Z_N^{3i} Z_N^{3j\star}) P_R\right)\chi^0_j Z_{\mu}
\end{eqnarray}
The interactions of Higgs boson with neutralino and chargino are given by
\begin{eqnarray}
&& {\frac{e}{2s_Wc_W}} \bar{\chi}^0_i \left[(Z_R^{1k} Z_N^{3j} -
Z_R^{2k} Z_N^{4j})(Z_N^{1i} s_W - Z_N^{2i} c_W)P_L\right.
\nonumber\\
&& + \left.(Z_R^{1k} Z_N^{3i\star} - Z_R^{2k}
Z_N^{4i\star}) (Z_N^{1j\star} s_W - Z_N^{2j\star} c_W) P_R\right]
\chi^0_j H^0_k \nonumber\\
&&  - {\frac{e}{\sqrt{2}s_W}} \bar{\chi}_i \left[(Z_R^{1k} Z_-^{2i}
Z_+^{1j} + Z_R^{2k} Z_-^{1i} Z_+^{2j}) P_L \right.  \nonumber\\
&& \left. + (Z_R^{1k} Z_-^{2j\star}
Z_+^{1i\star} + Z_R^{2k} Z_-^{1j\star} Z_+^{2i\star}) P_R\right]
\chi_j H^0_k
\end{eqnarray}
We know from the above equations that the pure bino does not interact with gauge boson or Higgs boson.
So the bino LSP annihilation can only proceed through coannihilation or mixing with higgsino or wino.
Thus the bino dark matter annihilation cross section is smaller than higgsino and wino dark matter.

\item[(ii)] \textbf{\textit{A}-mediated Case}:
In this case the LSP dark matter annihilates through the $s$-channel resonance of the CP-odd Higgs boson $A$,
usually called "$A$-funnel" \cite{Martin:1997ns}. Here we decouple all sfermions (fix them to 10 TeV)
and consider the resonance of $A$ which enhances the annihilation of dark matter.
The relavent interactions for the $A$-funnel annihilation processes are given by
\begin{eqnarray}
& & \frac{ie}{\sqrt{2}s_W} \bar{\chi}_{i} \left[(Z_H^{1k} Z_-^{2i}
Z_+^{1j} + Z_H^{2k} Z_-^{1i} Z_+^{2j}) P_L - (Z_H^{1k} Z_-^{2j\star}
Z_+^{1i\star} + Z_H^{2k} Z_-^{1j\star} Z_+^{2i\star}) P_R \right] \chi_j A^0_k
\nonumber \\
& & - {\frac{ie}{2s_Wc_W}} \bar{\chi}^0_i \left[(Z_H^{1k} Z_N^{3j} -
Z_H^{2k} Z_N^{4j}) (Z_N^{1i} s_W - Z_N^{2i} c_W) P_L \right. \nonumber \\
& & - \left. (Z_H^{1k} Z_N^{3i\star} - Z_H^{2k}
Z_N^{4i\star}) (Z_N^{1j\star} s_W - Z_N^{2j\star} c_W) P_R\right]
\chi^0_j A^0_k
\end{eqnarray}

\item[(iii)] \textbf{Coannihilation Cases}:
When a squark or stop has a mass approaching the LSP, it coannihilates with the LSP and helps to achieve
the correct relic density.
In both cases (squark coannihilation and stop coannihilation) we fix $A$ mass at 10 TeV.
For squark (stop) coannihilation, the squark (stop) mass is required to within 120$\%$ of
the LSP mass. The relevant interactions for squark coannihilation are given by
\begin{eqnarray}
&& U^-_i \bar{\chi}^0_1 \left[ \left( {\frac{-e}{\sqrt{2}s_Wc_W}}
Z_U^{Ii\star} (\frac{1}{3}Z_N^{11} s_W + Z_N^{21} c_W) - Y_u^I
Z_U^{(I+3)1\star} Z_N^{41} \right) P_L\right. \nonumber\\
&& + \left. \left( {\frac{2e\sqrt{2}}{3c_W}}
Z_U^{(I+3)i\star} Z_N^{11\star} - Y_u^I Z_U^{Ii\star} Z_N^{41\star}
\right) P_R \right] u^I + \mathrm{H.c.} \nonumber\\
&&+ D^+_i \bar{\chi}^0_1 \left[ \left( {\frac{-e}{\sqrt{2}s_Wc_W}}
    Z_D^{Ii} (\frac{1}{3}Z_N^{11} s_W - Z_N^{21} c_W) + Y_d^I
    Z_D^{(I+3)i} Z_N^{31} \right) P_L\right. \nonumber\\
&& + \left. \left( {\frac{-e\sqrt{2}}{3c_W}}
Z_D^{(I+3)i} Z_N^{11\star} + Y_d^I Z_D^{Ii} Z_N^{31\star} \right)
P_R\right] d^I + \mathrm{H.c.}
\end{eqnarray}
\end{itemize}

\begin{figure}[h]
\centering
\hspace*{-1.4cm}
\includegraphics[width=8.0in]{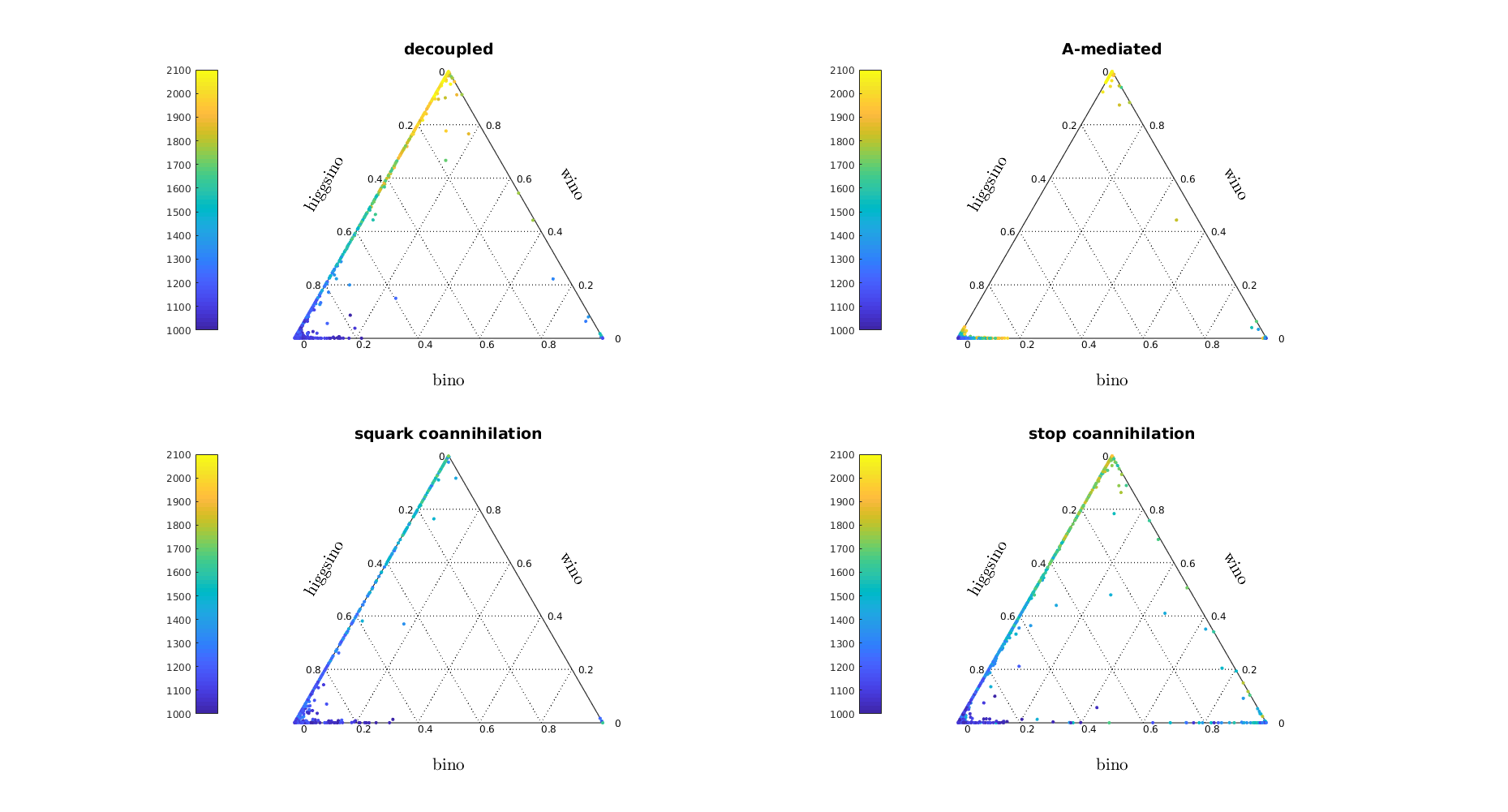}
\vspace{-1.5cm}
\caption{Scatter plots of the TeV neutralino dark matter showing the components
under the requirement of thermal relic density. The colors denote the dark matter mass
from 1 TeV to 2.1 TeV.}
\label{fig:components}
\end{figure}
In Fig. \ref{fig:components} we show the components of the TeV neutralino dark matter under the requirement
of correct thermal relic density. We see that in all four cases the  TeV neutralino dark matter is dominated
by higgsino or wino in order to satisfy the relic density.
Actually, a  pure higgsino (wino) dark matter around 1.1 (2.1) TeV has been found to give the required
thermal relic density in the literature \cite{Cohen:2013ama, Fan:2013faa, Bramante:2015una}.
The number of bino-like samples is quite small in the decoupled scenario, but increases
in the A-mediation and squark coannihilation cases, as expected.

\section{Constraints from dark matter detections}\label{section3}
\begin{figure}[h]
\centering
\includegraphics[width=6.5in]{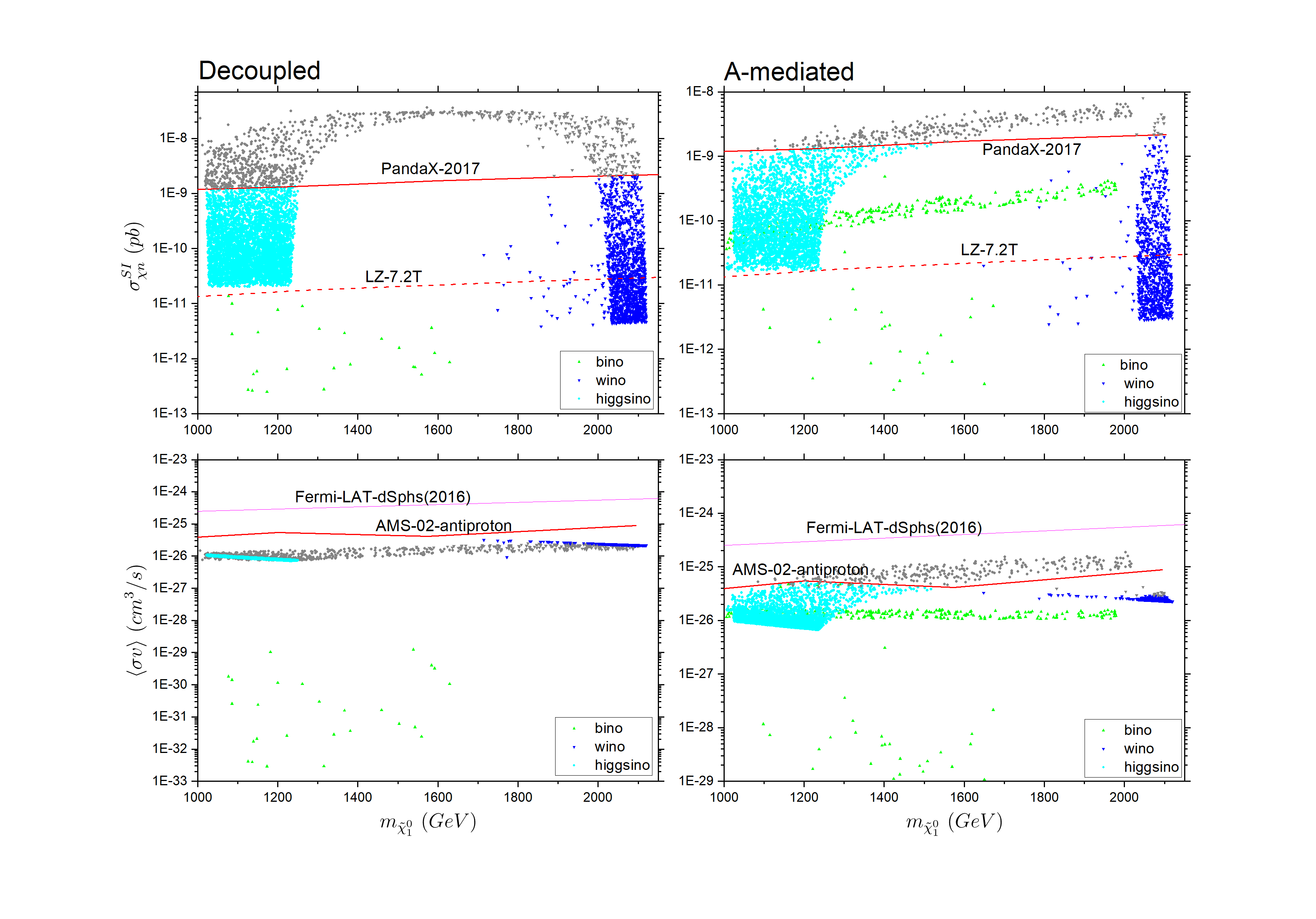}
\vspace{-2.1cm}
\caption{Scatter plots of parameter space satisfying the relic density at $2\sigma$ level.
The left panels are for the decoupled case while the right panels are for the \textit{A}-mediated case.
The upper parts show the spin-independent neutralino LSP-nucleon scattering cross sections
where the curves are the 90$ \% $ CL upper limits from PandaX (2017) \cite{Cui:2017nnn} and
the future sensitivity from LZ7.2T.
The lower parts show the dark matter annihilation cross sections $ \langle \sigma v \rangle $ where
the curves are the 95$ \% $ CL upper limits from the AMS-02 anti-proton data \cite{Cuoco:2017iax}
and the Fermi-LAT $\gamma$-ray data (dwarf spheroidal galaxies) \cite{Fermi-LAT:2016uux}.
The bino, wino and higgsino samples represent bino-like, wino-like and higgsino-like LSP, respectively.}
\label{fig:decoupled}
\end{figure}

The on-going direct and indirect detection experiments have constrained the dark matter interactions with the
standard model (SM) particles.
In Fig. \ref{fig:decoupled}, the upper panels show the spin-independent neutralino-nucleon
scattering cross sections where the current upper limits from PandaX and the future sensitivity
of LZ7.2T are plotted.
We see that the current PandaX data has excluded the region where higgsino and wino are mixed.
The future LZ7.2T experiment can cover the whole higgsino region and a major part of the wino region.

The lower panels of Fig. \ref{fig:decoupled} show the neutralino dark matter annihilation cross
sections where the upper limits from the AMS-02 anti-proton data
and the Fermi-LAT $\gamma$-ray data \cite{Fermi-LAT:2016uux} from
the observation of dwarf spheroidal galaxies
are plotted.
We see that the upper limits from these indirect detection data are weaker than the
direct detection limits (the parameter space above the AMS-02 anti-proton limits has already been
excluded by the PandaX data).

\begin{figure}[h]
\centering
\includegraphics[width=6.5in]{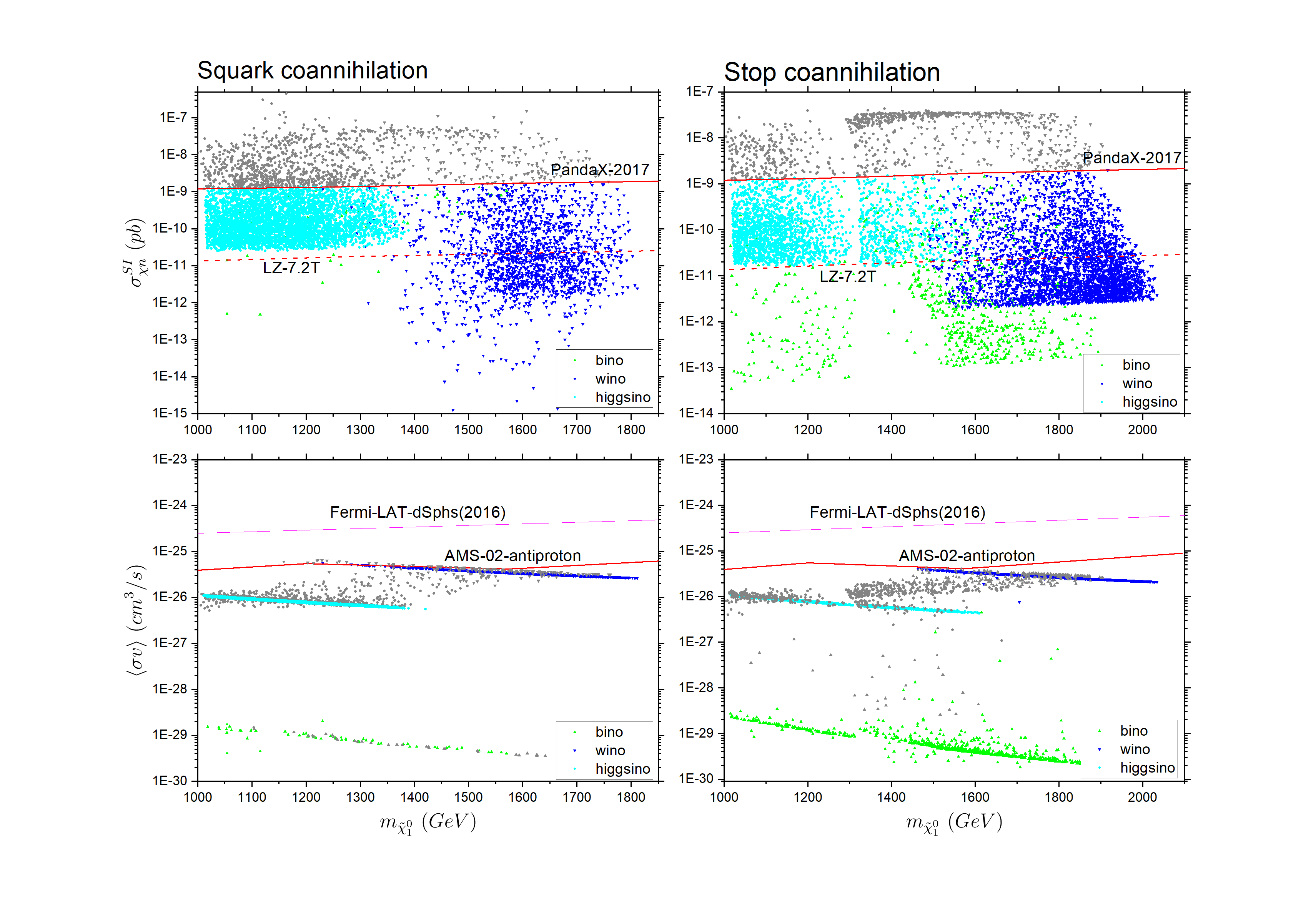}
\vspace{-2.1cm}
\caption{Same as Fig. \ref{fig:decoupled}, but for squark and stop coannihilation cases. }
\label{fig:coannihilation}
\end{figure}

In our calculation of the electron/positron and antiproton flux from the dark matter halo,
we changed the energy spectrum
of dark matter source term in GALPROP \cite{Strong:1998fr}. This source term is
\begin{align}
q_{\bar p, e^\pm}^{DM}(\textbf{x}, E_{kin}) = \frac{1}{2} {\left( \frac{\rho(\textbf{x})}{m_{DM}} \right)}^2 \sum_f \langle \sigma v \rangle_f \frac{dN_{\bar p, e^\pm}^f}{dE_{kin}}
\label{eq:sourece}
\end{align}
where $\bar{p}, e^\pm$ are antiproton and electron/positron, $\rho(\textbf{x})$ is the dark matter density
distribution, $m_{DM}$ is the dark matter mass, $\langle \sigma v \rangle_f$ is thermally averaged cross section
for dark matter annihilation into the SM final state $f$ (DM + DM $\rightarrow f\bar f$),
$dN_{\bar p, e^\pm}^f / dE_{kin}$ are the antiproton and eletron/positron energy spectrum per annihilation,
and the factor 1/2 is for the Majorana dark matter fermion. For $\rho(\textbf{x})$ we use NFW dark matter
density profile \cite{Navarro:1995iw}
\begin{align}
\rho_{NFW}(r) = \rho_0 \frac{r_0}{r} \left(\frac{r_0}{r_0+r} \right)^2
\end{align}
with the halo radius 20 kpc, the local dark matter density $\rho_0$ = 0.43GeV/cm$^3$ \cite{Salucci:2010qr}
at the solar position $r_0$ = 8kpc.

\begin{table}[th]
\caption{Four benchmark points, one point for each case (decoupled case, \textit{A}-mediated case, squark and stop
coannihilation cases) with largest $\langle \sigma v \rangle$.}
 \vspace*{0.5cm}
\begin{tabular}{*{11}{|c}|}
\hline
\multirow{2}{*}{$m_{DM}$} & \multirow{2}{*}{$\langle \sigma v \rangle$} & \multicolumn{5}{|c|}{Final states ($f \bar f$)} & \multicolumn{4}{|c|}{DM components}\\
\cline{3-11}
&&$W^+W^-$ & $Z^+Z^-$ & $\tau^+ \tau^-$ & $b \bar{b}$ & $t\bar{t}$ & $Z_N^{11}$ & $Z_N^{12}$ & $Z_N^{13}$ & $Z_N^{14}$\\
\hline
1463.6 &4.61 $\times$ 10\^(-26) &0.8730 &0.1260 &0      &0      &0      &0.0012 &-0.9978 &0.0557 &-0.0361\\
\hline
994.30 &1.36 $\times$ 10\^(-26) &0.6570 &0.2180 &0      &0      &0.0740 &0.2639 &-0.3153 &0.6531 &-0.6359\\
\hline
1714.9 &3.18 $\times$ 10\^(-26) &0.9999 &0      &0      &0      &0      &0.1369 &-0.99   &0.0312 &-0.0161\\
\hline
1267.1 &5.01 $\times$ 10\^(-26) &0.0720 &0.0530 &0.1150 &0.7300 &0.0139 &0.1619 &-0.0172 &0.6993 &-0.6960\\
\hline
\end{tabular}
\label{table}
\end{table}

\begin{figure}[h]
\centering
\includegraphics[width=6.5in]{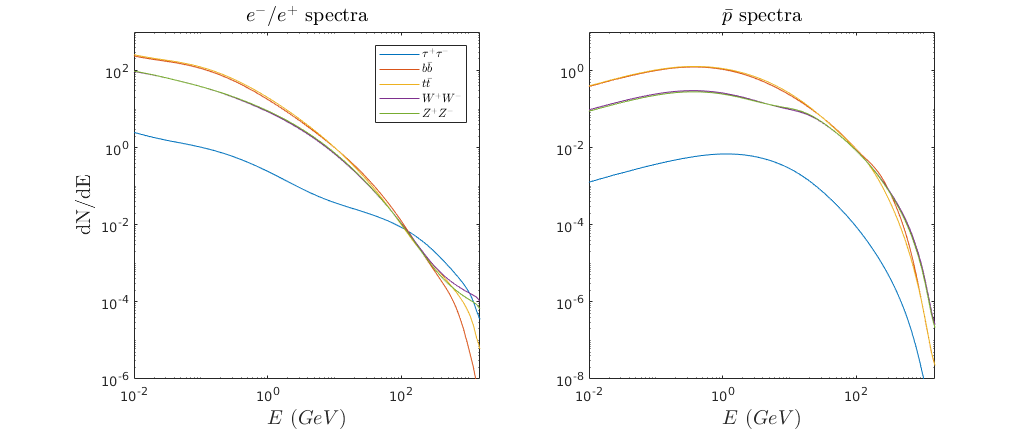}
\vspace{-1.1cm}
\caption{The positron/electron and antiproton spectra produced by the annihilation of two dark matter
particles with mass 1.5 TeV. }
\label{fig:spectra}
\end{figure}

From the source term we know that $m_{DM}$ and $\langle \sigma v \rangle_F$ are crucial for the intensity of the flux
and, therefore, we here choose one benchmark point for each case.
The energy spectrum $dN_{\bar p, e^\pm}^f / dE_{kin}$ data from \cite{Cirelli:2010xx} are used with the interpolation
method for the source term in GALPROP. In the 1-2 TeV dark matter mass range, the spectra of $e^\pm,~ \bar p$ are similar and
thus we only show the spectra for $m_{DM}$ = 1.5 TeV in Fig. \ref{fig:spectra}.
Also, we display in  Table \ref{table} four benchmark points, one point for each case with
largest $\langle \sigma v \rangle$.

\begin{figure}[h]
\centering
\includegraphics[width=6.5in]{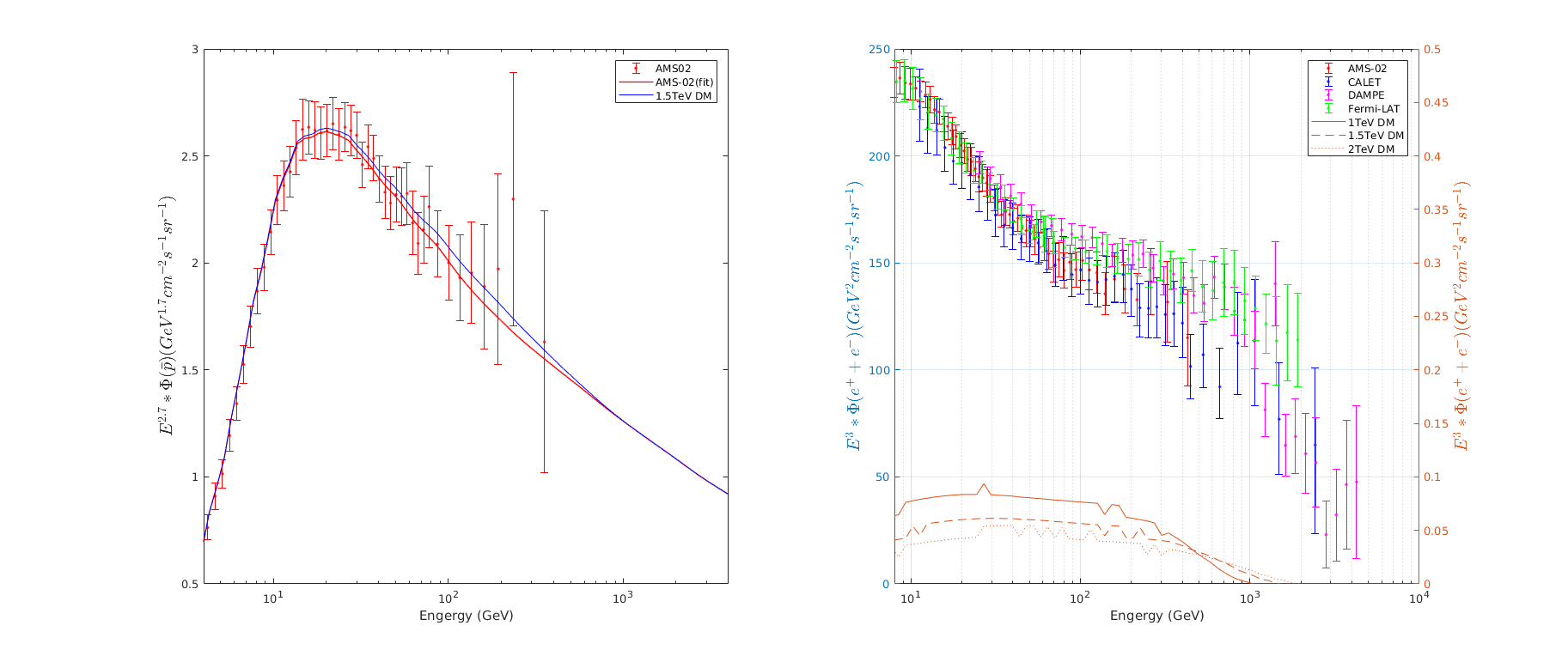}
\vspace{-1.1cm}
\caption{The antiproton flux (left panel) and electron plus positron flux (right panel).
For the antiproton flux, the AMS-02 data  \cite{Aguilar:2016kjl}
and its fitted curve as well as the results
with the 1.5 TeV neutralino dark matter annihilation contribution are shown.
For the electron plus positron flux, the AMS-02 \cite{Aguilar:2014mma}, the CALET \cite{Adriani:2017efm},
the DMAPE \cite{Ambrosi:2017wek} and the Fermi-LAT \cite{Abdollahi:2017nat} data
(read the left Y-axis) as well as the 1-2 TeV neutralino dark matter annihilation contribution
(read the right Y-axis) are shown.  The errors of the data are $1\sigma$ statistical and systematic. }
\label{fig:flux}
\end{figure}

Fig. \ref{fig:flux} is the antiproton and electron plus positron flux calculated by GALPROP,
compared with the experimental data. Here we see that the contributions of the 1-2 TeV neutralino
dark matter annihilation are too small to cause visible excess. This means that the plausible
electron/positron cosmic-ray excess at TeV energy reported by DAMPE \cite{Ambrosi:2017wek}
is not likely from the TeV neutralino dark matter annihilation. If this excess is verified, it may
point to some TeV leptophilic dark matter \cite{Duan:2017pkq}. However, the small contribution
to the antiproton flux from the TeV neutralino dark matter annihilation is still favored because so far
no excess has been observed for the antiproton flux.

In summary, the 1-2 TeV neutralino dark matter with correct thermal relic
density has been stringently constrained by the direct detection data. The constraints
of indirect detections from cosmic-ray flux are much weaker than direct detection limits.
The survived parameter space can be mostly covered by the future direct detection experiment
LZ7.2T. At the colliders, the TeV neutralino dark matter is hard to probe at the LHC \cite{Han:2013usa},
but can be effectively probed at a 100 TeV hadron collider.
For example, for a luminosity of 3000 fb$^{-1}$, a
100 TeV hadron collider can give a good probe for a TeV higgsino in the decoupled case \cite{Low:2014cba}
and 1-3 TeV for stop coannihilation case \cite{Cohen:2014hxa}.

\section{conclusion}\label{section4}
In this work we examined the thermal neutralino dark matter in a mass range of 1-2 TeV.
We considered various scenarios and confronted them with the latest
direct and indirect detections from PandaX and AMS-02/DAMPE.
We observed that the parameter space is stringently constrained by the direct detection limits.
In the allowed parameter space, the 1-2 TeV neutralino dark matter
annihilation contribution to the anti-proton flux is found to agree with the AMS-02 data while its contribution to
eletron/positron flux is too small to cause any visible excess.
The current survived parameter space can be mostly covered by a future direct detection experiment
LZ7.2T.

\section*{Acknowledgement}
This work was supported by the National Natural Science Foundation of China (NNSFC)
under grant Nos. 11705093,  11305049, 11675242 and 11375001,
by Peng-Huan-Wu Theoretical Physics Innovation Center (11747601),
by the CAS Center for Excellence in
Particle Physics (CCEPP), by the CAS Key Research Program of Frontier Sciences and
by a Key R\&D Program of Ministry of Science and Technology of China
under number 2017YFA0402200-04.

\bibliography{refs}

\end{document}